# Dark modes governed by translational-symmetry-protected bound states in the continuum in symmetric dimer lattices

Yixiao Gao,[1,2,3,*] Junyang Ge,[1,2,3] Shengzhi Sun,[1,2,3] and Xiang Shen[1,2,3]

[1]Laboratory of Infrared Materials and Devices, Research Institute of Advanced Technologies, Ningbo University, Ningbo 315211, China
[2]Key Laboratory of Photoelectric Detection Materials and Devices of Zhejiang Province, Ningbo 315211, China
[3]Engineering Research Center for Advanced Infrared Photoelectric Materials and Devices of Zhejiang Province, Ningbo University, Ningbo 315211, China
*Corresponding author: gaoyixiao@nbu.edu.cn

*Abstract* – Creating nonradiating dark modes is key to achieving high-$Q$ resonance in dielectric open cavities. The concept of photonic bound states in the continuum (BIC) offers an efficient method to suppress radiative loss through symmetry engineering. Structural reflection symmetry (RS) has been widely utilized to construct BICs in asymmetric metasurfaces. In this paper, we show that the radiation channel of translational-symmetry (TS) protected BIC in 1D symmetric dimer lattice could be unlocked by dimer spacing perturbation. A semi-analytical coupled mode analysis reveals that the total radiation suppression of the TS-BIC is due to the elimination of the first Fourier harmonic component in the lattice parameters. TS-BIC mechanism could also be applied in a 2D symmetric dimer lattice, and BICs protected by TS are robust to RS breaking, and vice versa, providing a promising way to independently control the quality factor of two interacting BIC resonances. Our results suggest a new degree of freedom to engineer BICs as well as their interactions in dimer lattices tailored by different symmetries, and could provide new insight for realizing practical applications requiring high-$Q$ resonances.

## 1. Introduction

Confining light using cavities is extremely important in modern science. The strongly enhanced light–matter interaction in a cavity allows the efficient implementations of numerous applications ranging from harmonic generation, small lasers, to ultrasensitive sensing [1]. The cavity quality factor $Q$ characterizes the strength of light-matter interaction, which is usually limited by material absorption and radiation losses. Radiative loss can be efficiently suppressed with the help of *dark modes* in open resonators to achieve high $Q$ resonance, especially for dielectric systems. Originated from quantum mechanics, bound state in the continuum (BIC) with a potentially infinite lifetime represents a general phenomenon in wave physics [2]. Photonic BIC is usually formed in an open system by engineering the radiation channels canceling out each other via destructive interference, offering an effective route to constructing dark mode. An ideal BIC resonance can be identified by the divergence of the $Q$ factor with complete radiation suppression, which is extremely difficult to externally excite and control. Researchers usually engineer cavity structure to deliberately introduce a leaky channel with a controlled radiation loss, and a high $Q$ factor could still be maintained in the form of a quasi-BIC (Q-BIC) resonance. Q-BICs has been widely explored in numerous applications requiring high-$Q$ and narrow-linewidth features [3-7].

Although various mechanisms have been exploited to produce BICs in optical systems [8-13], symmetry-protected BICs (SP-BIC) in metasurfaces attracts extensive research interests.

The structural symmetry of the unit cell is usually regarded as a key factor determining the formation of SP-BIC [14]. Typically, an SP-BIC resonance with an anti-symmetric modes profile would emerge in a symmetric structure due to its zero-coupling efficiency to the symmetric plane wave radiation [15]. An in-plane [4, 10, 16-18] or out-of-plane [19] structural perturbation that breaks reflection symmetry (RS) of the unit cell could introduce an net dipole moment in the unit cell, and thus unlock radiation channels to transform these SP-BICs into their quasi states. Structural asymmetry with broken RS will essentially make these Q-BICs polarization-sensitive. Recently, a symmetric dimer metasurface with a unit cell of two identical silicon disk was proposed to achieve Q-BIC resonance by controlling the dimer distance without breaking RS, and a multipolar analysis indicates an excitation of toroidal dipole (TD) resonance is responsible for the high-Q quasi-BIC resonance [20]. Further, a symmetric dielectric metasurface with four same cuboids in a unit cell was found possessing polarization-independent feature in the regime of BIC, and dual-band BIC resonances were associated with magnetic dipole (MD) and toroidal dipole (TD) resonances [21]. It is interesting that in these works BIC resonances is sensitive to the spacing perturbation within the unit cell, and a nonzero net dipole moment presents in unit cell even at the BIC resonance, which is quite different from asymmetric metasurfaces with RS breaking [10]. This raises the question — what is the underlying physics behind the BIC formation in these metasurfaces consisting of symmetric unit cells.

In this paper, we start with studying the resonant properties of the eigenmodes supported by a 1D symmetric dimer lattice, as well as their transmission characteristics upon the lattice parameters tuning towards the formation of BIC. Employing a semi-analytical coupled mode theory, we reveal the general mechanism governing the BIC resonance in a symmetric coupled dimer lattice and its transition into Q-BIC by translational-symmetry (TS) breaking through dimer spacing perturbation. We further investigate the BICs protected by TS and RS in 2D symmetric dimer lattice, show the different properties of two kinds of BICs against dimer gap and tilt angle perturbations, and numerically demonstrate an extreme Huygens metasurface [22] by the interaction between TS-BIC and RS-BIC.

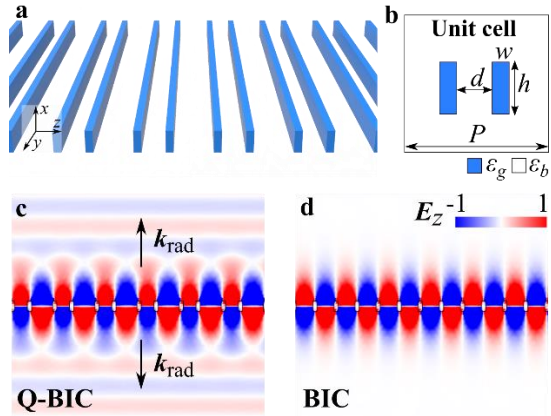

Fig. 1 (a) Schematic of the 1D dielectric dimer lattice. (b) Parameters of the dimer unit cell consisting of a pair of same dielectric bars, the blue region is the dielectric bar with a permittivity of $\varepsilon_g$ and the white surrounding medium has a permittivity of $\varepsilon_b$. (c) and (d) shows the electric field of the resonance in the dimer lattice at Q-BIC and BIC, respectively. The color is saturated for better illustration of the radiation field.

## 2. Results and discussions

Figure 1 shows the 1D dimer lattice consisting of identical dielectric bars that extend infinitely along the y axis, and are distributed periodically along the z axis. The bar dimer is located in the center of the unit cell with the height and width of bars as $h$ and $w$, the dimer spacing is $d$, and the period of the dimer lattice is $P$, as depicted in Fig. 1(b). The filling factor $\rho$ is defined as $\rho = 2w/P$. Dielectric bars have a permittivity of $\varepsilon_g$, and the whole lattice is embedded in a medium with permittivity $\varepsilon_b$. At a critical dimer spacing satisfying $d = P/2 - \rho P/2$, the dimer lattice is reduced to a uniform single bar lattice with a period of $P/2$, and we refer this critical dimer spacing $d$ as $d_c$ in what follows. The lattice is invariant by a translation of $P/2$ along the $z$-axis. We could define a translation operator **T** and the permittivity distribution of the lattice satisfies $\mathbf{T}\varepsilon(r + P/2) = \varepsilon(r)$. When $d$ is deviated from $d_c$, the original translational symmetry is broken [23] with $\mathbf{T}\varepsilon(r + P/2) \neq \varepsilon(r)$, and the lattice period returns to $P$. In this paper, without loss of generality, we consider the dielectric medium of the lattice is silicon with a refractive index of 3.5, and lattice is surrounded by homogenous medium with a refractive index of 1.

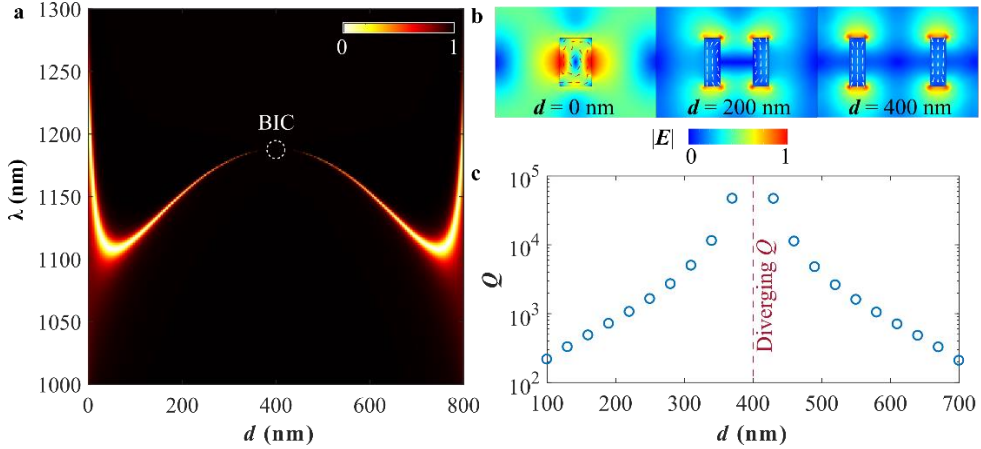

Fig. 2 (a) Transmission spectra of the dimer lattice with a varying spacing $d$. BIC state is highlighted by dashed circle. (b) Electric field profile in unit cell of the dimer lattice. (c) Quality factor $Q$ as a function of spacing $d$ of the dimer lattice. The vertical dashed line indicates the quality factor is diverging at $d = 400$ nm.

Next, we show the spacing-$d$-dependent transmission spectra in the dimer lattice. We consider a normally incident (along $x$ axis) $z$-polarized plane wave excitation and the parameters of dimer lattice are $w = 100$ nm, $h = 300$ nm, and $P = 1000$ nm, and thus $d_c = 400$ nm. Figure 2(a) shows the evolution of the transmission spectra upon the increasing dimer spacing $d$ from 0 nm to 800 nm. When $d = 0$ nm, the dimer lattice becomes a single bar lattice with a bar width of 200 nm. A resonance dip appears at 1228 nm with its the electric field profile depicted in Fig. 2(b), and a vortex distribution of electric field (the red arrows) inside the dielectric bar indicate that the lattice mode is related to a magnetic dipolar resonance [24, 25]. With the separation $d$ increasing, the spectral position of the resonance initially experiences a blue shift when $d$ is less than 50 nm, and then redshift to a longer wavelength before $d$ reaching 400 nm. During this process, the resonance linewidth monotonously decreases and the electric profiles share the similarity of vortex distribution, for example, as can be seen from Fig. 2(b) depicting the field distribution at $d = 200$ nm. The corresponding resonance $Q$ factor is plotted in Fig. 2(c), we could also see an increasing trend of $Q$ with $d$ from 100 nm to 400 nm, which is consistent with the inverse quadratic law between $Q$ factor and perturbation parameter in quasi-BIC resonances reported in Ref. [10]. At the critical dimer spacing $d_c = 400$ nm, the transmission dip disappears, as denoted by the dashed circle, and the $Q$ factor also shows a diverging nature, as the red dashed line in Fig. 2(c) depicted, indicating the formation of a completely dark BIC resonance in the dimer lattice with $d = 400$ nm. We found the BIC resonance is located at $\lambda = 1189.6$ nm through an eigenfrequency calculation, the mode profile of which is depicted in Fig. 2(b). In Fig.1(d), we illustrate the mode profile at BIC in a larger region, and a clear suppression of radiation could be observed compared with Fig.1(c) where a Q-BIC resonance is presented in the dimer lattice with $d = 300$ nm. It is worth noting that the radiating electric fields above and below the dimer lattice have a π phase difference at Q-BIC. With $d$ further increase from 400 nm to 800 nm, the variation of transmission spectra shows an opposite trend with respect to $d$ from 0 nm to 400 nm, and the resonant wavelength and width are identical with the same $\Delta d = |d - d_c|$ value. In this range, the dielectric bars start to couple with neighboring unit cells, and the overall result is the same lattice is shift by $P/2$ in the $x$-direction.

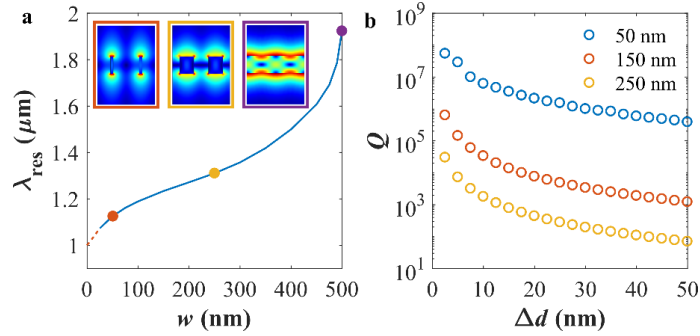

Fig. 3 (a) Resonant wavelength $\lambda_{res}$ of the dimer lattice in BIC regime as a function of $w$. Insets depict the field profiles of the coupled lattice with $w$ equaling to 50 nm (red), 250 nm (yellow) and 500 nm (purple). (b) The quality factor $Q$ of the resonance as a function of spacing perturbation $\Delta d$ in the dimer lattice with different width $w$ equal to 50 nm, 150 nm and 250 nm.

The dielectric bar width $w$ of the dimer lattice has a significant effect on the BIC resonance characteristics. Figure 3(a) shows the influence of $w$ on the resonant wavelength at BIC state of the dimer lattice. The resonant wavelength monotonically increases with the width $w$. When $w$ is equal to $P/2$ (i.e. $w = 500$ nm in our case), the dimer lattice becomes a slab waveguide with a thickness of 300 nm, the calculated resonant wavelength is 1928.7 nm. The resonance mode profile is depicted in the inset with purple frame of Fig. 3(a), which is equivalent to the interference pattern of two counter propagating fundamental TM modes with effective mode indices of 1.928. While at a smaller $w$ value (e.g. $w = 50$ nm), the BIC resonant field mainly localized near the dielectric bars and the evanescent field extends into the ambient medium with a longer distance. When the decreasing $w$ approaches to 0, the BIC wavelength reaches towards 1 μm, i.e. the period of lattice, as indicated by the red dashed curve in Fig. 3(a). However, a thin dimer structure can hardly confine electromagnetic field, and thus the BIC state can no longer be sustained.

$Q$ factors of BIC resonances in the dimer lattice are sensitive to spacing perturbations. Figure 3(b) shows the dependence of $Q$ factor of Q-BIC state on the spacing perturbation $\Delta d$ defined as $d - d_c$ with different $w$. Similar with Fig. 2(c), the $Q$ factor drops with a larger $\Delta d$ for all $w$ values due to an enhanced radiation loss. However, $Q$ factor in a wider width is more sensitive to $\Delta d$. When $\Delta d = 0$ nm, all three $w$ cases are in BIC resonance with infinite $Q$ factor. With a spacing perturbation $\Delta d = 2.5$ nm and 50 nm, the $Q$ factor of a $w = 50$ nm dimer lattice keeps a high value at $5.4 \times 10^7$ and $3.8 \times 10^5$, respectively. While for a $w = 250$ nm dimer lattice, the $Q$ factor substantially decreases to $3.0 \times 10^4$ and 69.9 at $\Delta d = 2.5$ nm and 50 nm, respectively, indicating $Q$ factor of Q-BIC state in a thinner dimer lattice is more robust to $\Delta d$ perturbation, making them sensitive to possible fabrication imperfections in order to achieve a high-$Q$ resonance.

To gain a deeper insight into the physics of BICs formation, we employ a semi-analytical model to analyze the resonances in symmetric dimer lattice. Since we consider TM incidence, only the $y$-component of the magnetic field is nonzero, the magnetic field distribution of the resonance in the dimer lattice is governed by the Helmholtz equation [26]

$$\frac{\partial}{\partial x}\left(v\frac{\partial H_y}{\partial x}\right)+\frac{\partial}{\partial z}\left(v\frac{\partial H_y}{\partial z}\right)+k_0^2 H_y = 0 \tag{1}$$

where $v = 1/\varepsilon(x, z)$ is the inverse of the dielectric function, $k_0$ is the free space wavevector. $H_y$ field could be approximated by the summation of couter-propagating waves and radiation wave as

$$H_y(x,z) \approx (Ae^{iKz} + Be^{-iKz})\varphi(x) + H_{rad} \tag{2}$$

where $A$ and $B$ are slowly varying envelopes of two counter-propagating waves, $\varphi(x)$ is the transverse mode profile of the unmodulated slab, $H_{rad}$ is the radiation field, and $K = 2\pi/P$. $v(x, z)$ is expressed as a summation of Fourier components as

$$v(x,z) \approx v_0(x) + v_1(x)\cos(Kz) + v_2(x)\cos(2Kz) \tag{3}$$

where the Fourier components of $v$ in the lattice region are

$$\begin{aligned}v_0 &= \rho/\varepsilon_g + (1-\rho)/\varepsilon_b \\ v_1 &= -2\Delta\varepsilon/(\pi\varepsilon_g\varepsilon_b)\cos(\pi(2d+P\rho)/2P)\sin(\pi\rho/2) \\ v_2 &= -\Delta\varepsilon/(\pi\varepsilon_g\varepsilon_b)\cos(\pi(2d+P\rho)/P)\sin(\pi\rho)\end{aligned} \tag{4}$$

and for regions outside the dimer lattice, $v_0 = 1$, and $v_n = 0$. In the studied configuration, $A$ and $B$ are equal in Eq. (2) [26]. Substituting Eqs. (2) and (3) into Eq. (1), we obtain the surface normal radiation wave $H_{rad}$ is governed by

$$\frac{\partial}{\partial x}\left(v_0\frac{\partial}{\partial x}\right)H_{rad} + k_0^2 H_{rad} = -A\frac{\partial}{\partial x}\left(v_1\frac{\partial}{\partial x}\right)\phi(x) \tag{5}$$

The radiation field $H_{rad}$ could be solved through Green's function [26, 27], and can be expressed as

$$H_{rad} = -v_1 A\int_0^h G(x,x')\frac{\partial^2}{\partial x^2}\phi(x)dx' \tag{6}$$

From Eq. (6) we could see the amplitude of radiation field is proportional to the first order Fourier component $v_1$, when the critical dimer spacing $d_c = 400$ nm is reached, $v_1$ equals to zero, leading to a total suppression of radiation field and thus a BIC resonance formed [28]. We should point out that any spacing perturbation $\Delta d$ will break the translational symmetry of lattice on BIC resonance causing a period doubling and the emergence of first order Fourier component of the dielectric modulation.

This BIC formation by the Fourier component engineering could also be applied to a 2D dimer lattice. Figure 4(a) illustrates the schematic of a 2D dimer square lattice. Here, we consider the lattice thickness $h = 300$ nm, the lengths of each cube in a unit cell along $z$ and $y$ axes are $w = 150$ nm and $l = 400$ nm, the periods along $z$ and $y$ have the same value as $P = 1000$ nm, and the spacing $d$ of the dimers varies. The dimer in the unit cell is invariant under 180° rotation around the $x$ axis ($C_2$ operation), also has a reflection symmetry with respect to $xOz$ (i.e. invariant under $\sigma_y$ operation) and $xOy$ (i.e. invariant under $\sigma_z$ operation) plane. The unit cell has a $C_{2v}$ symmetry [29]. For the 2D dimer lattice, the corresponding material parameter (i.e. the inverse of permittivity $v$) could be expanded as a summation of 2D Fourier components as $v = v_{0,0} + \Sigma_{(m,n)\neq(0,0)} v_{m,n} \exp(imK_z z + inK_y y)$ with $K_z = K_y = 2\pi/P$ [28]. The 2D dimer lattice is excited by normally incident $z$-polarized plane wave. Figure 4(b) shows the evolution of transmission spectra of the symmetric dimer lattice upon the varying $d$. A similar trend as Fig. 2(a) could be observed: at a relatively small $d$, e.g. $d = 100$ nm, the Q-BIC resonance is located at 1140 nm with a $Q$ factor of 116. With $d$ continuously increasing to 350 nm, the transmission dip experiences linewidth narrowing and red-shifting. At a critical dimer spacing at $d_c = 350$ nm, a dark BIC state emerges in the lattice with a 1167 nm resonant wavelength and a diverging $Q$. At this critical spacing, the dimer lattice is equivalent to a single cube lattice with a period of $P/2$ in $z$-direction, leading to the elimination of $v_{1,n}$ and thus a radiation suppression. These Q-BIC modes are also closely related to MD lattice modes. As depicted in Fig. 4(c), a typical MD resonance field profile is presented at $d = 0$ nm. While for $d = 200$ nm, the vectorial electric field (white arrows) of Q-BIC mode also has a vortex distribution, indicating an effective magnetic dipole moment along the $y$-axis is presented in the gap region.

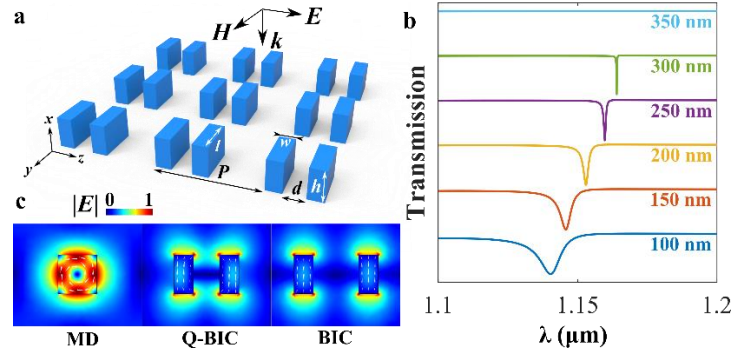

Fig. 4 (a) Schematic of the 2D dimer lattice. (b) Transmission spectra of the 2D dimer lattice as a function of $d$ ranging from 100 nm to 350 nm. (c) Electric field profile on $z$-$x$ plane in the center of a dimer at MD ($d = 0$ nm), Q-BIC ($d = 200$ nm) and BIC ($d = 350$ nm) resonances.

At the critical dimer spacing $d_c$, the 2D dimer lattice support two kinds of BICs, and Fig. 5(a) shows the normalized surface charge density $\sigma$ of two types of BIC resonance: the left one is the BIC considered in this paper, which is protected by TS and possesses a MD moment along $y$ axis constructed by a pair of out-of-plane antiphased electric dipoles, and the right one is the case reported in Ref. [10], which is protected by RS and has a MD moment along $x$ axis constructed by a pair of in-plane antiphased electric dipoles, and a resonant wavelength at 1254 nm. Figure 5(b) illustrates the spacing (left) and tilt angle (right) perturbations that breaks the TS and RS of the dimer unit cell. Here, TS breaking is referred to as a dimer spacing perturbation to $d_c$, leading to the period of dimer lattice along z axis returned to $P$, and RS breaking is referred to as the unit cell being no longer invariant under a $\sigma_y$ operation. We first analyze the multipolar components of the TS- and RS-protected BIC resonances with varying perturbation strength [20], to reveal the underlying physics of the radiation channels opening in quasi-BICs in the 2D dimer lattice. Figure 5(c) shows the electric and magnetic dipolar moments in the TS-BIC resonance in a unit cell. We could find that the TS-BIC resonance possesses a dominant magnetic dipolar moment along $y$-axis with the varying perturbation parameter $\Delta d$. When $\Delta d = 0$ nm, i.e. a dimer lattice without TS perturbation, the TS-BIC resonance also possess a nonzero magnetic dipole moment. While for the BIC resonance protected by RS, with the growing strength of RS perturbation $\Delta\theta$ from 0° to 20°, an electrical dipole moment aligning with $z$-axis (also with incidence polarization) emerges and increase with $\Delta\theta$, which is in stark contrast with that of TS-BIC. It is noted that an out-of-plane magnetic dipole moment $m_x$ is also presented in the RS-BIC resonance, while it does not contribute to the radiation channel. Figure 5(e) shows the magnetic field distribution of TS-BIC resonance and its quasi-states with $\Delta d = 100$ nm in six periods of a 2D dimer lattice. We could observe that the dimer lattice becomes a series of out-of-plane electric dipoles with their dipole moments alternating their direction along $z$-axis, as could be inferred from the surface charge density $\sigma$ distribution in a dimer in Fig. 5(a), and two neighboring anti-phased electric dipoles could form a magnetic dipole. For the TS-BIC resonance, the synthesized magnetic dipoles between the evenly-spaced electric dipoles have the same magnitude but are antiphased to their neighbors, leading to a perfectly destructive interference with total suppression of radiation. However, for the perturbed TS-BIC resonance with nonzero $\Delta d$, the synthesized magnetic dipoles would have a larger magnitude between two closer electric dipoles, and the radiation channel is open due to the failure of perfect destructive interference between neighboring antiphased magnetic dipoles with unequal magnitudes, which is typically a lattice effect related to the first order Fourier component of the dimer lattice parameter as discussed above.

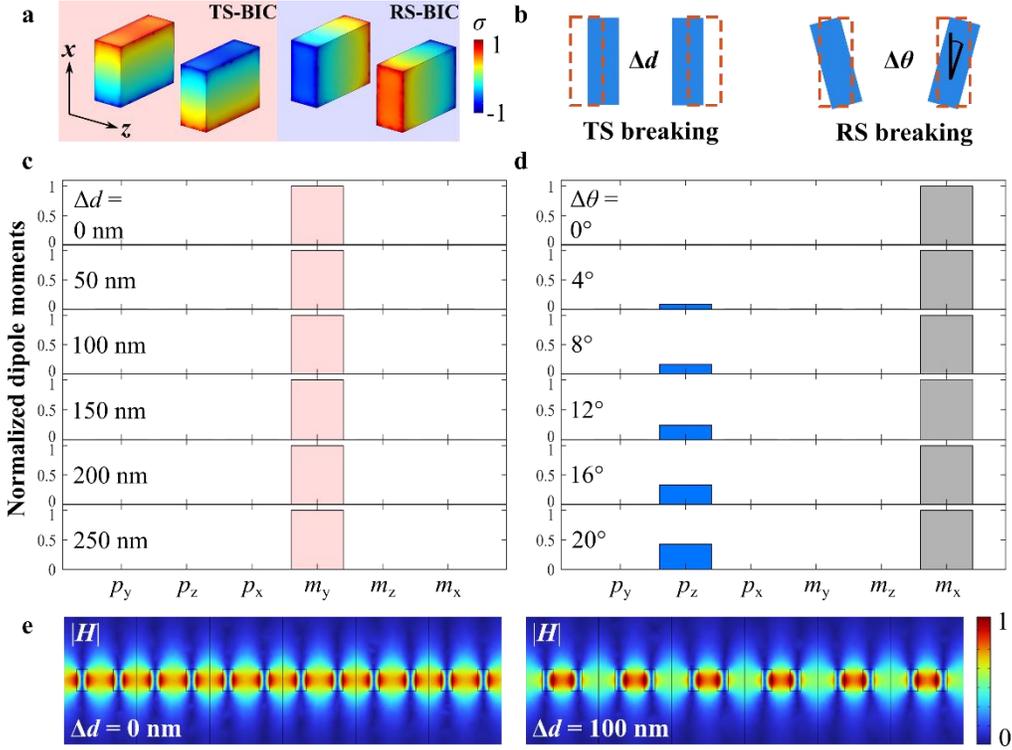

Fig. 5 (a) Normalized surface charge density of TS-BIC and RS-BIC states at $d_c$ = 350 nm. (b) Illustration of dimer spacing and tilt angle perturbation that breaks TS and RS in a dimer unit cell, and red dashed frame indicates a dimer working in the regime of BIC. (c) Normalized dipole moments in TS-BIC resonance with dimer spacing perturbation $\Delta d$ varying from 0 nm to 250 nm. (d) Normalized dipole moments in RS-BIC resonance with tilt angle perturbation $\Delta\theta$ varying from 0 deg to 20 deg. (e) Magnetic field distribution of TS-BIC resonance with $\Delta d$ = 0 nm (left) and 100 nm (right).

Figure 6(a) shows the transmission spectra of dimer lattices with different $\Delta\theta$ perturbations. Here, we set the dimer spacing $d$ = 250 nm to ensure the TS-BIC working in its quasi-state. As expected, the RS-BIC located at 1246 nm wavelength is sensitive to the $\Delta\theta$ perturbation, exhibiting a growing resonance linewidth with a larger $\Delta\theta$ [10]. While the TS-QBIC is robust to the $\Delta\theta$ perturbation: the resonant wavelength maintains at ~ 1163 nm, and a limited influence on $Q$ factor could be observed. Therefore, we could conclude that TS-BIC is robust to RS perturbation, and vice versa, due to different BIC formation mechanisms relating to the radiation suppression. This feature offers a flexible route to independently tuning the $Q$ factor of BIC resonances by perturbing different symmetries. We note here that a more general discussion on BIC protected by different symmetries based on group theory as well as their application in multifunctional nonlocal devices are reported in Refs. [30, 31].

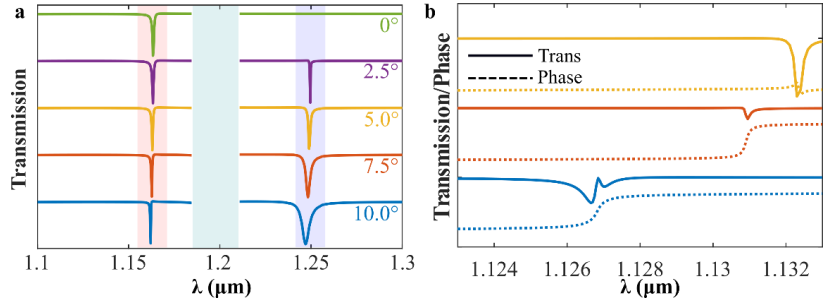

Fig. 6(a) Transmission spectra of a dimer lattice with tilt angle $\Delta\theta$ perturbations. (b) Transmission spectra (solid) and phase shift (dashed) of perturbed dimer lattice with varying $d$, the dimer parameter is tuned to ensure TS-BIC and RS-BIC having the same resonant wavelength.

A pair of spectrally overlapped degenerate resonances with opposite parity could construct Huygens metasurface with $2\pi$ phase coverage and low back reflection, which could be applied in phase modulators. Electric and magnetic dipole resonances with broad linewidths are usually employed to fulfill this requirement. Pure phase modulation could only be achieved with the help of tunable medium exhibiting a large refractive index change, such as liquid crystal [32]. Overlapping two high Q factor quasi-BIC resonances could realize so called extreme Huygens metasurface with an extreme phase dispersion within a narrow frequency range, which is highly suitable for realizing electro-optically phase modulation because of a small refractive index change on the order of $10^{-2}$ or less driven by electro-optical effects [33]. The opposite field parity along *x*-axis of TS-BIC and RS-BIC could further be explored to construct extreme Huygens' metasurface [22] for pure phase modulation [34]. Avoiding transmission fluctuation in the extreme Huygens metasurface is key to facilitate a phase modulation capability with high transmission efficiency. Typically, similar Q factors of two interacting quasi-BIC resonances would ensure a nearly flat transmission spectrum. The independent Q factor manipulation of TS-QBIC and RS-QBIC would provide an effective method to matching the Q factors of two interacting QBIC resonances. Table 1 shows the resonant wavelengths and Q factors of TS-BIC and RS-BIC with different dimer spacing when $l = 290.56$ nm, $\Delta\theta = 3.5°$ and other parameters are the same as above. We could see that the two QBICs has a similar resonant wavelength but TS-QBIC has a larger variation of $Q$ factor, while the $Q$ factor of RS-QBIC is relatively robust against $\Delta d$ perturbation. Figure 6(b) shows the transmission spectra of perturbed dimer lattice with different $d$. When $d = 250$ nm, due to large $Q$ factor difference between two QBICs, the transmission ratio at the $2\pi$ phase jump remains in a relatively low value (the lowest ratio is 0.63) in a wide wavelength range. The transmission at the $2\pi$ phase jump is improved due to a similar $Q$ factor of two BICs at $d = 300$ nm, the lowest ratio is 0.84, indicating the formation of extreme Huygens' metasurface with high transmission efficiency. It is noted that a finer tuning the parameters of the dimer unit cell could further improve the transmission efficiency. When $d = 350$ nm, the TS-BIC returns to its dark state which is inaccessible by the free-space radiation, and the transmission spectrum exhibits a single resonant dip with a $\pi$ phase shift owing to the excited RS-QBIC resonance.

**Table 1 Resonant wavelengths and $Q$ factors of TS-QBIC and RS-QBIC**

| | TS-QBIC | | RS-QBIC | |
| --- | --- | --- | --- | --- |
| $d$ (nm) | $\lambda_{res}$ (nm) | $Q$ factor | $\lambda_{res}$ (nm) | $Q$ factor |
| 250 | 1126.9 | 1738 | 1126.8 | 6040 |
| 300 | 1130.9 | 7126 | 1130.9 | 6911 |
| 350 | 1132.3 | $4.68\times10^5$ | 1132.3 | 6741 |

It should be noted that the features of TS-BIC and RS-BIC resonances could also be explored in Q factor sensitive applications. In the perfect absorption realized by degenerate critical coupling [35], the absorption rates of two overlapped resonances are usually different due to different resonance field profiles. Independently tuning the radiative rates of QBIC resonances through different symmetry perturbations would simultaneously match the radiative Q factor and nonradiative Q factor of each resonance, leading to a unity absorption ratio at the resonant wavelength, while this condition is difficult to be satisfied for QBIC resonances protected by the same symmetry [36].

For practical implementations, despite free-standing dielectric membranes could be utilized to harness BIC resonance [37], patterning the dielectric nanostructures on a substrate would provide a more flexible way to engineering the lattice unit cells and is more reliable for long term stability. However, a substrate (refractive index $n_{sub}$) may introduce additional leaky channels which could further degrade the $Q$ factors of Q-BICs [38]. Through tuning the lattice period $P$ as well as filling factor and ensuring the resonant wavelength of Q-BICs larger than $n_{sub}P$, the leaky channel in the substrate could be effectively suppressed so that BIC resonance could still be maintained in the dielectric lattices [21, 39]. It is also worth mentioning that the dimer lattices investigated here are all assumed to be infinite, a practical finite sized dielectric lattices would has a lower $Q$ factor compared with their infinite counterparts, nevertheless, a relatively high-Q could still be obtained as long as the lattice is large enough, for example, Ref. [39] reported a Q-BIC resonance with a $Q$ factor of 18511 in a 27×27 silicon block array.

## 3. Conclusions

In conclusion, we investigate dark modes governed by BIC in a symmetric dimer lattice, by breaking the translational symmetry through deviating the dimer spacing $d$ from its critical value $d_c$, the BIC resonance collapse into a Q-BIC with radiation controlled by the strength of $d$ perturbation. For a dimer lattice with thinner $w$, the Q-BIC resonance is more robust against $d$ perturbation in terms of Q factor. We employ a semi-analytical coupled mode theory to reveal the underlying physics behind the formation of BIC in the symmetric dimer lattice, and find that the surface normal radiation wave is proportional to the first order Fourier component $\nu_1$ of the dielectric parameter (i.e. the inverse of dimer lattice permittivity $\nu$) modulation, and a dimer lattice with $d = d_c$ lead to a vanishing $\nu_1$ and thus total radiation suppression. The BIC formation mechanism could also be applied to 2D dimer lattices, and the TS-BIC in the dimer lattice is insensitive to tilt angle perturbations. Through carefully tuning the perturbation influencing the $Q$ factor, an extreme Huygens' metasurface could be achieved by coupling TS-BIC and RS-BIC in the same dimer lattice. Our result suggests TS protected BICs could offers a new degree of freedom for engineering high-Q resonance as well as resonance coupling for nanolasing, sensing and nonlinear applications.

## Acknowledgment

This work was supported by National Natural Science Foundation of China (Grant No. 62105172), Zhejiang Provincial Natural Science Foundation of China (Grant No. LQ21F050004), Ningbo Natural Science Foundation (Grant No. 202003N4102), The K. C. Wong Magna Fund in Ningbo University.

**Declaration of Interests**: The authors declare no conflicts of interest.